\documentstyle[12pt]{article}
 
 \def\ds{\displaystyle}
\def\BB{\dot{A}}
\def\etal{{\em et al.}}

\def\beq{\begin{equation}}
\def\eeq{\end{equation}}
\def\be{\begin{equation}}
\def\ee{\end{equation}}
\def\bea{\begin{eqnarray}}
\def\eea{\end{eqnarray}}

\def\MAKESPACE{\phantom{\int_{\int_{A_B}}}}

\newcommand{\eqref}[1]{eq.~(\ref{#1})}   
\def\newline{\hfill\break}
\def\VEV#1{\left\langle#1\right\rangle}
\def\qbq#1{\bar{#1}#1}

\catcode`\@=11 
\def\lsim{\mathrel{\mathpalette\@versim<}}
\def\gsim{\mathrel{\mathpalette\@versim>}}
\def\@versim#1#2{\vcenter{\offinterlineskip
        \ialign{$\m@th#1\hfil##\hfil$\crcr#2\crcr\sim\crcr } }}
\catcode`\@=12 

\begin{document}

\begin{titlepage}
\begin{flushright}
CERN-TH/96-320\\
IASSNS-HEP-96/114\\
TAUP-2388-96\\
WIS-41/96-Nov\\
hep-ph/9611466\\
\end{flushright}

\vfill
\begin{centering}
{\large{\bf
The $\Delta I ={1\over2}$ Rule in the Light of Two-Dimensional QCD\\
}}
\vspace{.5cm}
{\bf John Ellis}\\
\vspace{.05in}
Theoretical Physics Division, CERN, CH-1211 Geneva 23, Switzerland \\
e-mail: john.ellis@cern.ch \\
\vspace{0.5cm}
{\bf Yitzhak Frishman}\\
\vspace{.05in}
Department of Particle Physics \\
Weizmann Institute of Science \\
Rehovot 76100, Israel\\
e-mail: fnfrishm@wicc.weizmann.ac.il\\
\vspace{0.5cm}
{\bf Amihay Hanany}\\
IAS, Princeton 08540, NJ, USA\\
e-mail: hanany@sns.ias.edu\\
\vspace{0.2cm}
and\\
\vspace{0.2cm}
{\bf Marek Karliner}\\
\vspace{.05in}
School of Physics and Astronomy
\\ Raymond and Beverly Sackler Faculty of Exact Sciences
\\ Tel-Aviv University, 69978 Tel-Aviv, Israel
\\ e-mail: marek@vm.tau.ac.il \\
\vspace{0.6cm}
{\bf Abstract} \\
\bigskip
{\small
We calculate in QCD$_2$ the ratios of baryonic matrix elements of 
$\Delta I = 2$ and $\Delta I = 0$ four-fermion
operators, with a view to understanding better the mechanism of
$\Delta I = 1/2$ enhancement in QCD$_4$. We find relatively small
suppressions of both the scalar-scalar and vector-vector $\Delta I = 2$
four-fermion operators. We discuss the possible implications of these
results, in view of a suggestion that gluon condensation may be an
important contributing factor in the $\Delta I = 1/2$ enhancement seen
in QCD$_4$. At the technical level, our calculation of the vector-vector
operator matrix element requires a treatment of the time dependence of the
QCD$_2$ soliton which had not been developed in previous phenomenological
calculations within this model.
} 
\end{centering}
\vfill
\begin{flushleft}
CERN-TH-96/320\\
November 1996 \hfill
\end{flushleft}
\end{titlepage}
\vfill\eject

\section{Introduction}
Two-dimensional QCD (QCD$_2$) is often a useful testing ground for
ideas concerning non-perturbative effects in QCD in four dimensions
(QCD$_4$). Among the issues studied in QCD$_2$ have been those
associated with confinement~\cite{thooft}, the appearance of 
baryons ($B$) as solitons~\cite{FS},
the existence of constituent quarks and
their relation to current quarks~\cite{quarks}, 
and quark-antiquark
condensates~\cite{FK}.
However, care must be exercised in carrying conclusions over directly
from QCD$_2$ to QCD$_4$, since there are essential differences between
the dynamics in the different dimensions.
Among these differences are the dimensionless nature of the QCD$_4$
gauge coupling as opposed to the dimensional nature of the
QCD$_2$ gauge coupling, the associated differences in the
infra-red behaviours of QCD$_2$ and QCD$_4$ (at least in perturbation
theory), and the absence in QCD$_2$ of spontaneous chiral symmetry
breaking and the pseudo-Goldstone boson interaction of the light
pions in QCD$_4$. Another important difference, to which we shall
return in this paper, is the fact that gluons in QCD$_2$ do not
have any physical polarization states.
Thus in 2D the analogue of
the glue condensate in the vacuum $\langle0|F^2|0\rangle $ 
may only come from mixing with quark-antiquark pairs, and 
hence is expected to be much smaller than in QCD$_4$.
In particular, it vanishes in the large-$N_c$ limit.

QCD$_2$ has been used previously to calculate certain non-perturbative
matrix elements of local operators, such as quark-antiquark condensates
in baryons $\langle B | \bar{q} q | B \rangle $~\cite{FK}, with 
results~\footnote{Of particular interest have been
ratios of quark condensates, such as the ratio of the strange
quark condensate to the sum of all condensates, for which
a smooth non-zero limit exists in the limit of vanishing
quark masses.}
that compare well with Skyrme models~\cite{DN} and phenomenological
determinations in
QCD$_4$~\cite{ChengLi}, and may
cast light on observations of apparent violations of
Okubo-Zweig-Iizuka (OZI) rule in meson-baryon
couplings~\cite{Erice}.
Non-leptonic weak decays are controlled by the matrix elements
of higher-dimensional operators, such as
$\langle B | ( \bar{q} q ) ( \bar{q} q ) | B \rangle$, in
QCD$_4$. These matrix elements exhibit some puzzling features,
in particular large enhancements of $\Delta I = {1\over2}$ transitions
compared with $\Delta I = {3\over2}$ transitions (for a review,
see for example \cite{Cheng}).
It seems that
calculable perturbative QCD$_4$ renormalization effects below
the weak scale~\cite{pert} cannot explain all the large $\Delta
I={1\over2}$
enhancement observed, and hence that the major part of it must
be non-perturbative in nature. The main purpose of this paper
is to calculate an analogue in QCD$_2$ of the enhancement
of $\Delta I = {1\over2}$ component of
non-leptonic decay amplitudes, in the hope of casting light on the
nature of the non-perturbative enhancement mechanism active in
QCD$_4$.
 
This calculation requires certain improvements in the technology
developed previously for calculations in QCD$_2$~\cite{FS}.
As is well known, baryons in QCD may be described as solitons
in a bosonized formulation. The collective coordinates of the
soliton must be quantized consistently, introducing a time
dependence into the soliton wavefunction which corresponds to the
angular momentum of the baryonic soliton (Skyrmion) in
QCD$_4$~\cite{Skyrme}.
Previous calculations of baryonic matrix elements of simple local
operators such as $\bar{q} q$~\cite{FK} did not require an explicit
treatment
of this collective-coordinate quantization. However, this is necessary
for the calculation of the baryonic matrix element of the non-leptonic
weak Hamiltonian, which contains current-current terms
$J^\mu\, J_\mu$, since the time component $J^0$ involves time
derivatives of the bosonic fields in the soliton.
 
Our main physical result is a relatively small numerical factor
between the non-perturbative $\Delta I={1\over2}$ and
$\Delta I={3\over2}$ matrix elements, which appears inadequate
to explain the $\Delta I={1\over2}$ enhancement observed in
QCD$_4$, even allowing for the perturbative enhancement factors
calculated in QCD$_4$~\cite{pert}. We interpret this result as indicating
that
the observed $\Delta I={1\over2}$ enhancement in QCD$_4$
may be due to particular features of the non-perturbative QCD$_4$
dynamics that are absent in QCD$_2$. A candidate for this is the
gluon condensate, which has indeed been proposed~\cite{glue} as the main
mechanism responsible for the observed $\Delta I ={1\over2}$ enhancement
in QCD$_4$, and which is relatively small in two dimensions,
even vanishing in the large-$N_c$ limit.

It is interesting to compare this result to previous calculations in
four-dimensional Skyrme models~\cite{Skyrmehalf}, which found a
magnitude of $\Delta I = 1/2$ enhancement that is sensitive to the
model-dependent spatial wave function. In our two-dimensional case, 
the integral over this wave function can be carried out analytically
and exactly. The Skyrme models incorporate correctly the physics of
quark condensation and chiral symmetry breaking in four
dimensions, but do not in general include representations of
the gluonic degrees of freedom in QCD$_4$, and in particular the
physics of gluon condensation.
 
\section{Computation of the Matrix Elements for the Quadrilinear
Scalar Interaction}
 
As a warm-up exercise, we start in
this section by calculating an analogue of the
$\Delta I = 1/2$ enhancement in the ratio between matrix
elements of four-fermion scalar densities in QCD$_2$.
Our eventual aim is to calculate the analogous
ratio between matrix elements of products of
Lorentz-vector currents, as
this is more comparable to the quantity of interest in
four dimensions. However the latter calculation involves more
technical aspects of baryonic solitons in QCD$_2$, so to
get the flavor of the physics of the
computation we first tackle the simpler case of scalar densities
with $I_3 = 0$ and $I = 2, 0$, which we denote by
$T_{20}$ and $T_{00}$ respectively.
 
To calculate the matrix elements of these operators, we first
review the static classical soliton which describes a baryon in
QCD$_2$~\cite{FS}. In the strong-coupling limit
this is given by a solution to a Sine-Gordon equation, as seen in
 eq.~(5.4) of Ref.~\cite{FS}, namely
\beq
\varphi(x)={4\over \beta}\tan^{-1} [\exp{\beta\, \sqrt{2}\,m\,x}]
\label{phiDef}
\eeq
where $\beta=\sqrt{4\pi\over N_c}$ is
the coupling constant of the Sine-Gordon theory,
${m\over\beta}$ is the mass of the soliton, and $m$ is
related to the common
bare mass of the quarks by a renormalization group
relation appropriate to two dimensions.
The semiclassical quantization of this soliton entails the
introduction of time-dependent coordinates
$z_i,\quad i=1\ldots N_f$ \cite{FS}.
In terms of these coordinates, the relations between
${\bar q} q$ bilinear densities and the scalar fields follow from
the standard bosonization correspondence:
\beq
\qbq{q_i} = \left\{ 1 + (\cos\beta\varphi-1) |z_i|^2 \right\} \Lambda
\label{qbqFromg}
\eeq
where $\Lambda$ is an appropriate mass scale \cite{FK}.
 
On physical grounds, and in
order to calculate finite terms in the classical expression
for the four-fermion operators, we need
to subtract the vacuum expectation values of the operators.
This leads to the relations
\bea
&&\qbq{q_i}\qbq{q_j}
-\VEV{0|\qbq{q_i}\qbq{q_j}|0}=\hskip20em\nonumber\\
&=&\left\{\left[1 + \left( \cos \beta \varphi - 1 \right)|z_i|^2\right]
  \left[1 + \left( \cos \beta \varphi - 1 \right)|z_j|^2\right]
-1\right\}\,\Lambda^2 \label{qbqqbq}\\
&=&
\left\{\left( \cos \beta \varphi - 1 \right)\left(|z_i|^2+|z_j|^2\right)
+\left( \cos \beta \varphi - 1 \right)^2 |z_i|^2|z_j|^2\right\}
\,\Lambda^2
\nonumber
\eea
where the term proportional to
${-}1$ at the end of first row comes from the vacuum subtraction.
 
Next we turn to the computation of the expectation values
(\ref{qbqqbq}). In the
semiclassical approximation, we may separate out the
classical contribution, which amounts to an integral over the spatial
coordinate $x$. Setting $\beta\sqrt{2} m = \mu$, and using some
elementary $x$ integrals listed in the Appendix, we find
\beq
\VEV{\qbq{q_i}\qbq{q_j}}=
\left[{16\over3}|z_i|^2|z_j|^2
-4\left(|z_i|^2+|z_j|^2\right)\right]{\Lambda^2\over\mu}
\label{VEVqbqqbq}
\eeq
where the notation $\langle\ldots\rangle$ denotes the
integration of equation~(\ref{qbqqbq}) over $x$.
 
In order to average over the quantum coordinates $z_i$,
we must select a
suitable wave function for the baryon. Since direct analogues
of the lowest-lying baryon octet do not exist in QCD$_2$,
we choose the
$\Delta^{+}$ state, which is the closest analogue of the proton in
two dimensions\footnote{We note in passing that the $\Delta I = 1/2$
rule is known experimentally
to be valid for $\Omega^- \rightarrow \Xi \pi$ decays in QCD$_4$: the ratio
of $\Delta I = 3/2, 1/2$ amplitudes $A_{3,1}$ is measured to be
$A_3 / A_1 = -0.063 \pm 0.014$ \cite{Bourquin}.}.
Its wave function is proportional to $z_1^2\,z_2$, and since we are
interested in ratios of matrix elements, the normalization of the state
factors out. Our convention is to normalize the $z$ integral 
(\ref{genericintegral})
to unity 
for $N=1,P=2$ when $N_f = N_c = 3$, as is appropriate for the physical
$\Delta^+$ state.
 
We now discuss in more detail the internal symmetry properties of
the bilinear ${\bar q_i} q_j$ and
quadrilinear ${\bar q_i} q_j {\bar q_k} q_{\ell}$
operators. The isospin-one bilinear scalar densities
are given in terms of the $u$ and $d$ fields by
\be
\begin{array}{lcc}
|1\,,\phantom{{-}}1\rangle & =  &\phantom{{-}}(\bar d u)  \\
&&\\
|1\,, {-}1 \rangle & =  &{-}(\bar u d)   \\
&&\\
|1\,,\phantom{{-}} 0 \rangle & =  &\phantom{{-}}
{\displaystyle (\bar u u) - (\bar d d)\over\displaystyle\sqrt{2} }\\
\end{array}
\ee
The corresponding relations between the $I_3 = 0$
quadrilinear scalar densities of isospins $I = 2, 0$
and the quark fields are
\be
\begin{array}{ll}
 T_{20} & =  \VEV{
\ds  \sqrt{{2\over 3}}
 \left[{\ds(\bar u u) - (\bar d d)\over\ds\sqrt 2}\right]^2 -
{2\over\sqrt 6}(\bar u d)(\bar d u) }\\
\\
& = \VEV{
{-}{\ds 1\over\ds\sqrt{6} }
\left\{2(\bar u d)(\bar d u) - \left[(\bar u u) - (\bar d d)
\right]^2\right\}}
\end{array}
\ee
 
\be
\begin{array}{ll}
T_{00} & = \VEV{
{-} \ds{1\over\sqrt 3}\left\{2(\bar u d)(\bar d u) +\ds \left[{(\bar u u ) -
 (\bar d
d)\over\sqrt 2}\right]^2\right\} }\\
\\
& = \VEV{
\ds {-}{1\over 2\sqrt 3}\left\{4(\bar u d)(\bar du) + [(\bar u u) - (\bar d
d)]^2\right\}}
\end{array}
\ee
%
We then use (\ref{IcIIx}) to perform the $x$ integrations,
obtaining
\bea
T_{20} &\propto&  \left[2|z_1|^2|z_2|^2 - (|z_1|^2 - |z_2|^2)^2 \right]
\ds{\Lambda^2\over\mu}
=\left[4z_1|^2|z_2|^2 - (|z_1|^4 + |z_2|^4) \right]
\ds{\Lambda^2\over\mu}\nonumber\\
\label{T20T00}\\
 T_{00}&\propto&  {4|z_1|^2|z_2|^2 + (|z_1|^2 -
 |z_2|^2)^2\over\sqrt{2}} \,\,\ds {\Lambda^2\over\mu}
={2|z_1|^2|z_2|^2 + (|z_1|^4 + |z_2|^4)\over\sqrt{2}}\,\,
\ds {\Lambda^2\over\mu}
\nonumber
\eea
for the different quadrilinear scalar densities with $I = 2,0$,
respectively.
 
The ratio of interest to us is
\be
{\langle T_{20}\rangle\over \langle T_{00}\rangle} =
\sqrt 2  \,{2R - 1\over R + 1}
\label{intermsofR}
\ee
where the auxiliary ratio $R$ is defined by
\be
R\equiv 2\,{\langle|z_1|^2z_2|^2 \rangle\over
\langle|z_1|^4 + |z_2|^4 \rangle}
\label{Rratio}
\ee
Using the integrals (\ref{genericintegral}), we get
$R = {2\over 3}$, which leads to the final result
\be
{\langle T_{20}\rangle  \over \langle T_{00}\rangle } = {\sqrt 2\over 5}
\approx 0.28.
\label{scalarratio}
\ee
for the ratio of quadrilinear scalar densities.
 
This ratio is not nearly as small as the corresponding ratio
for current-current operators in QCD$_4$~\cite{Bourquin}, even after
allowing for Clebsch-Gordan coefficient factors of order unity and
the calculable perturbative short-distance enhancement in
the latter case~\cite{pert}. However, before concluding anything about
the significance of different non-perturbative effects in
QCD$_2$ and QCD$_4$, we must analyze the technically more
complicated case of bilinear products of
Lorentz-vector currents in QCD$_2$.
 
\section{Formalism for Vector Currents and Collective-Coordinate 
Quantization}
 
The extra complications in the calculations of hadronic matrix elements of
Lorentz-vector
currents and their bilinear products arise
because one must take into account the time dependence
of the hadronic wave function~\cite{FS}, an issue that has not been
confronted
in previous QCD$_2$ calculations~\cite{FK}, including those in the
previous section. These complications appear
for the first time because the expressions for the vector currents
involve time derivatives of
the $z_i$, corresponding physically to the appearance of
conjugate momenta.
 
To see how these arise, we first derive general expressions
for the vector-current
operators in terms of the classical coordinate $x$ and
the quantum coordinates $z$.
The usual bosonization rules yield for the vector currents
\bea
J_+ & = {\ds \strut i\over\ds\strut 2\pi}\, U^\dagger\partial_+ U, \nonumber \\
&\\
J_- & = {\ds \strut i\over\ds\strut 2\pi}\, U \partial_- U^\dagger,\nonumber
\label{currents}
\eea
where $U$ is a nonlinear $\sigma$-model variable in the action of the
bosonized representation of QCD$_2$.
The semiclassical quantization rules give the field $U$ in terms of a
time-dependent ``zero-mode'' rotation
$A(t)$ from the classical solution $U_c(x)$:
\be
U = A(t)\,U_c (x)\,A^\dagger(t)
\label{rotation}
\ee
The last column of the zero-mode matrix $A$ is expressed
in terms of the semiclassical variables $z_i$.
 
The light-front derivatives $\partial_{+,-}U$ may now be
evaluated:
\be
\partial_{+,-}U  = \BB (t) U_c A^\dagger (t)
+A(t) U_c \BB^\dagger (t) \pm A(t) U_c'(x) A^\dagger (t)
\label{partials}
\ee
Using (\ref{partials}), we then find
\begin{eqnarray}
 U^\dagger\partial_+U & = & AU^\dagger_c(A^\dagger\BB) U_c A^\dagger
 + A\BB^\dagger
 + AU^\dagger_c U_c' A^\dagger   \nonumber \\
&& = A[(U^\dagger_c-1) + 1] A^\dagger
\BB [(U_c-1)+1]A^\dagger + A\BB^\dagger
x 4+A U_c^\dagger U_c'A^\dagger  \nonumber \\
&& =A(U^\dagger_c-1) A^\dagger
\BB (U_c-1)A^\dagger+A(U_c^\dagger-1)A^\dagger\BB A^\dagger
 \nonumber \\
&& +\BB (U_c - 1) A^\dagger + A U^\dagger_c U_c'A^\dagger
\end{eqnarray}
which yields the following
expression for the current $J_+$:
\begin{eqnarray}
J_+  =  {i\over 2\pi}
\left\{   A(U^\dagger_c {-} 1) A^\dagger
\BB (U_c {-} 1) A^\dagger
  +[\BB (U_c {-} 1) A^\dagger {-} A(U^\dagger_c {-} 1) \BB^\dagger]
 +A(U_c^\dagger U'_c)A^\dagger \right\} \nonumber\\
\label{Jplus}
\end{eqnarray}
and an analogous expression may be derived for $J_-$.
 
Defining
$\Phi \equiv \exp\left(i \beta \varphi\right) - 1$,
we may rewrite (\ref{Jplus}) in the form
\beq
(J_+)_{kl}  = {i\over 2\pi}
\left(z_k|\Phi|^2z_j^*\dot z_j z^*_l +\dot z_kz^*_l\Phi^*  -z_k\dot
 z^*_l\Phi
 -i\beta\varphi'(x) z_kz^*_l\right)
\label{Jphi}
\eeq
which may be simplified to
\beq
(J_+)_{kl}  =  {i\over 2\pi}\left[2\left(1-\cos\,\beta\,
\varphi\right) z_k (z_j^*\dot z_j) z^*_l
 -i\beta\,\varphi' z_kz^*_l
 +\dot z_k z^*_l \,\Phi
-z_k\dot z^*_l \,\Phi \right]
\label{finalJplus}
\eeq
The analogous calculation for the current $J_-$ yields
\beq
(J_-)_{lk} = {i\over 2\pi} \left[2\left(1-\cos\,\beta
\varphi\right)z_l (z_j^*\dot z_j)z_k^*
- i\beta\varphi 'z_lz^*_k
 +\dot z_lz^*_k\Phi^*
 -z_l\dot z^*_k\Phi^* \right]
\label{finalJminus}
\eeq
Equations (\ref{finalJplus}) and (\ref{finalJminus}) are
the basis for our subsequent discussion of the matrix
elements of the current-current operators.
 
It is convenient to rewrite (\ref{finalJminus}) and (\ref{finalJplus})
in terms of the `momentum' variables
\beq
\pi_l  \equiv  {1\over 2M} \left[\dot z_l - z_l(z^*_j\dot z_j)\right].
\label{momenta}
\eeq
which are convenient for computational purposes, though they differ from
the conjugate momenta resulting from the action by terms
linear in $z$, as seen below in equation~(\ref{newconstraint}).
We may rewrite the currents (\ref{finalJplus},\ref{finalJminus})
as follows in terms of the `momenta' (\ref{momenta}):
\begin{eqnarray}
(J_+)_{kl} & = & \left({i\over 2\pi}\right)\left\{\left(1-
\cos\beta\varphi\right)(2M)\left(z_k\pi^*_l - \pi_kz^*_l
\right) \right. \nonumber
 \\
&&-i\left(\sin\beta\varphi\right)(2M)\left(\pi_kz^*_l +
 z_k\pi^*_l\right)
\nonumber \\
& & \left. -i \beta\varphi 'z_kz^*_l \right\}
\label{currentsA}
\end{eqnarray}
\begin{eqnarray}
(J_-)_{lk}  & = & \left({i\over 2\pi}\right)\left\{\left(1-\cos
\beta\varphi\right)(2M)\left(z_l\pi^*_k - \pi_lz^*_k\right)
\right. \nonumber \\
&& +i\left(\sin\beta\varphi\right)(2M)\left(\pi_lz^*_k +
 z_l\pi^*_k\right)
\nonumber \\
&& \left. -i \beta\varphi 'z_lz^*_k\right\}
\label{currentsB}
\end{eqnarray}
and are ready to address the computation of the matrix elements.

We now formulate the quantization of
the quantum-mechanical soliton problem.
In terms of the $z$, $\dot{z}$ variables,
the bosonized effective action of QCD$_2$ may be written as \cite{FS}
\beq
S[z] = {1\over 2M}\int dt (Dz)^\dagger(Dz) + {i\over 2} N_c\int dt  [\dot
 z^\dagger\cdot z
- z^\dagger\cdot\dot z]
\label{action}
\eeq
where $ (Dz)_k  = \dot z_k - z_k (z^*_l\dot z_l)  $
is the covariant derivative with respect to the
auxiliary $U(1)$ gauge field.
The $z$ are subject to the constraints
\be
z^*\cdot z \, = \, 1, \,
z^*\cdot (Dz) \, = \, 0
\label{constraints}
\ee
where the first constraint follows from the $CP^{N-1}$ nature of the
$z_i$-s and
then the second follows from the definition of $D z$.
The canonical momenta are given by
\bea
p_k={\partial S\over \partial\dot z^*_k} = {1\over 2M} \left[\dot z_k - z_k
(z^*_l\dot z_l)\right] + {i\over 2} N_cz_k
\eea
and there is a conjugate expression for $p_k^*$.
It is convenient to rewrite the constraints
(\ref{constraints}) in the form
\beq
\begin{array}{ccl}
p_k &=& {1\over 2M} (Dz)_k + {i\over 2} N_c z_k \\
&&\\
z^* p &=& {i\over 2} N_c
\end{array}
\label{newconstraint}
\eeq
In order to quantize the system of coordinates with constraints, we
impose the commutation relations
\beq
i[p_k, z^*_l] = \delta_{kl} - z_kz_l^*
\eeq
corresponding to the following commutation rules for the
`momenta' $\pi_i$:
 \beq
[z^*_k, \pi_l] = i(\delta_{kl} - z^*_kz_l)
\label{quantization}
\eeq
which maintain consistency with the constraints
$z_k z_k^*=1$  and $z^*_l \pi_l=0$.
The first of these follows from 
the unitarity condition on the matrix $A$,
and the second  follows from the first, combined with the 
definition of $\pi_l$, eq.~(\ref{momenta}).

\noindent
We now represent $i \pi_l$ by
\beq
i \pi_l = 
{\partial \over  \partial z_l^*}- z_l z_k^* {\partial\over z_k^*}
\label{pilrep}
\eeq
This representation incorporates the commutation rules
eq.~(\ref{quantization})
and the constraint $z_l^* \pi_l =0$, given that $z_l^* z_l =1$.
Using (\ref{pilrep}), we find
\beq
i \pi_l f(z, z^*) = i [\pi_l, f(z, z^*)]  
\label{picomm}
\eeq
which will be used in the calculations of the following section.

\section{Computation of the Matrix Elements for the Vector
Current-Current Interaction}
 
We now apply the formalism developed above to the calculation of the
ratio of matrix elements in which we are interested.
As in the scalar case, we calculate the following ratio
of flavour combinations, where the Lorentz indices are left
implicit:
\beq
{T_{20}\over T_{00}} = \sqrt 2\,
{\langle[2(\bar u d)(\bar d u) - (\bar u u - \bar d
 d)^2] \rangle
\over
\langle[4(\bar u d)(\bar d u) + (\bar u u - \bar d d)^2]\rangle}
\label{vectorratio}
\eeq
The numerator and the denominator in (\ref{vectorratio}) involve
similar products of Lorentz components of the vector currents:
\beq
\begin{array}{lcl}
\langle(\bar u d)(\bar d u)\rangle &\propto &
\langle (J_+)_{12} (J_-)_{21} \rangle + \langle (J_-)_{21} (J_+)_{12} 
\rangle\,, \MAKESPACE \\
\langle(\bar u u)(\bar d d)\rangle &\propto &
\langle (J_+)_{11} (J_-)_{22} \rangle + \langle (J_-)_{11} (J_-)_{22}
\rangle
\end{array}
\label{components}
\eeq
and similarly for $\langle (\bar u u)(\bar u u)\rangle, \langle (\bar d d)
(\bar d d)\rangle$.
Recalling the expressions (\ref{currentsA}, \ref{currentsB})
for the components appearing in (\ref{components}), we see that
the evaluation of the matrix elements
requires the following commutation
relations equations derived from (\ref{quantization}):
\bea
\left\{\begin{array}{ccl}
i[\pi^*_1, z^2_1] &=& 2z_1 (1 - z_1^*z_1) \\
&&\\
i[\pi^*_1, z_2] &=& - z^*_1 z_2 \\
&&\\
i[\pi^*_1, z^2_1z_2] &=& z_1z_2 (2 - 3 z_1^*z_1) \\
&&\\
i[\pi^*_2, z^2_1z_2] &=& z^2_1(1 - 3z^*_2z_2)
\end{array}\right.
\eea
We will also need the results
\beq
i(z_1\pi^*_2 - \pi_1z^*_2) z^2_1z_2
 = z^3_1(1 - 2z^*_2z_2) 
\label{commone}
\eeq
and
\beq
i(\pi_1z^*_2 + z_1\pi^*_2)^2 z^2_1z_2
= z^3_1(1 - 4 z^*_2z_2)
\label{commtwo}
\eeq

Using these relations, it is possible to rewrite
$\VEV{(\bar{u}d)(\bar{d}u)}$ in the form
\beq
\langle (J_+)_{12} (J_-)_{21} + h.c. \rangle \propto
A_{11} \, + \, A_{22} \, + \, A_{33} \, + \, (A_{31} + A_{13})
\label{As}
\eeq
where the different terms appearing in (\ref{As}) may be expressed as
\bea
A_{11} & = & {2(2M)^2\over (2\pi)^2}\left[\int dx \left(1-\cos
\beta\varphi\right)^2\right]\int\mid z_1\mid^6(1 - 2\mid
z_2\mid^2)^2 \nonumber\MAKESPACE\\
A_{22} & = &  {-}\,{2(2M)^2\over (2\pi)^2} \left[\int dx\left(\sin
\beta\varphi\right)^2\right]\int\mid z_1\mid^6(1 - 4\mid
z_2\mid^2)^2 \nonumber\MAKESPACE\\
A_{33} & = &  {2\over (2\pi)^2}\left[\int
dx\left(\beta\varphi'\right)^2\right]\int\mid z_1\mid^6\mid
z_2\mid^4 \MAKESPACE\\
A_{31} + A_{13} & = & {2(2M)\over (2\pi)^2}(2\pi) \int\mid z_1\mid^6\mid
 z_2\mid^2
(1 - 2\mid z_2\mid^2)\nonumber
\label{Aexpressions}
\eea
These different terms come from products and combinations of the
three parts of (\ref{currentsA}) and (\ref{currentsB}). 
It is easy to see that the only interference terms that
survive the integral over
$x$ are the $A_{13,31}$ terms exhibited in (\ref{Aexpressions}).

Performing the $x$ integrals over the classical configuration
parametrized by $\varphi(x)$, and using
formulae given in the Appendix, we obtain
\bea
A_{11} & = &{2M^2\over \pi^2}{1\over\mu}\left({16\over 3}\right)\int\mid
z_1\mid^6\left(1-2\mid z_2\mid^2\right)^2 \nonumber\MAKESPACE\\
A_{22}&= &-{2M^2\over\pi^2} {1\over\mu}\left({8\over
3}\right)\int\mid z_1\mid^6(1-4\mid z_2\mid^2)^2 \nonumber\MAKESPACE\\
A_{33}&=  &{1\over 2\pi^2}(8\mu)\int\mid z_1\mid^6\mid z_2 \mid^4 
\MAKESPACE\\
(A_{31}+A_{13}) &= &
 {2M\over \pi^2}(2\pi)\int\mid z_1\mid^6\mid z_2\mid^2(1-2\mid
z_2\mid^2)\nonumber
\eea
Using the relation between the two mass scales $M$ and $\mu$
which is given in the Appendix, as well as the $z$ integrals
collected there, we obtain finally the numerical value
\beq
\langle  (\bar u d)(\bar d u)\rangle  = 
{\mu\over 7}\left[{10\over 27} +
\left({2\over\pi}\right)^2\right]
\label{ubddbu}
\eeq
for the first terms in the numerator and denominator of (\ref{vectorratio}).

Turning now to the remaining terms, we recall that
\bea
(J_\pm)_{kk} & = &\left({i\over 2\pi}\right)
\left\{\left(1 - \cos\beta\varphi\right)\left(z_k\pi_k^* -
\pi_kz^*_k \right)(2M) \right.  \nonumber\\
& \mp & i\left(\sin\beta\varphi\right)\left(\pi_kz^*_k +
z_k\pi_k^*\right)(2M)  \nonumber\\
& - &\left.  i\beta\varphi' z_kz^*_k\right\} \nonumber\\
& = & \left({i\over 2\pi}\right)\left\{(2M)\left(1-\cos\beta
\varphi\right)\left(z_k\pi^*_k - z_k^*\pi_k + i(1-z_kz^*_k)\right) \right.
 \nonumber\\
& \mp & i(2M)\left(\sin\beta\varphi\right)\left(z^*_k\pi_k +
 z_k\pi^*_k
- i(1 - z_kz^*_k)\right) \nonumber\\
 & - & \left. i\beta\varphi' z_kz^*_k\right\}
\eea
The commutation relations may then be used to evaluate
\bea
\begin{array}{lcl}
i(z_1\pi_1^* - z_2\pi^*_2) z_1^2z_2 \,
&=&z^2_1z_2(2-3z^*_1z_1) - z^2_1z_2(1-3z^*_2z_2)^2 \MAKESPACE\\
&=&z^2_1z_2[1-3(|z_1|^2 - |z_2|^2)] \\
\end{array} 
\label{z1pi1z2pi2}
\eea
Using this relation, we see that the action of the
currents on the proton state gives
\bea
\begin{array}{lcl}
[(J_\pm)_{11} - (J_\pm)_{22}] \,|p\rangle  
&=& \ds {1\over 2\pi}
\Bigl\{\,2M\left(1-\cos\beta\varphi\right)[1-2(|z_1|^2 -
|z_2|^2)]  \MAKESPACE\\
&&\phantom{aaaa}
\mp i \left(\sin\beta\varphi\right)(2M)[1-4(|z_1|^2 - |z_2|^2)]
\MAKESPACE \\
&&\phantom{aaaa}
+\beta\varphi'
(|z_1|^2 - |z_2|^2)\,\Bigr\}\, (z^2_1z_2)\, |g\rangle
\end{array}
\label{jproton}
\eea
which results in the amplitude
\bea
\begin{array}{l}
{\cal I}(x,z_1,z_2) \equiv
\langle p|\, [(J_+)_{11} - (J_+)_{22}][(J_-)_{11} 
- (J_-)_{22}]\,|p\rangle  
\,+\,\langle\, [\ldots]_{-}\,[\ldots]_{+}\,\rangle =
\MAKESPACE\\
=\ds {2\over(2\pi)^2} |z_1|^4|z_2|^2 \Bigl\{(2M)^2\left(1-\cos
\beta\varphi\right)^2[1-2(|z_1|^2 - |z_2|^2)]^2 \MAKESPACE\\
-(2M)^2\left(\sin\beta\varphi\right)^2\,[1-4(|z_1|^2 -
 |z_2|^2)]^2 \MAKESPACE\\
+\left({4\pi\over N_c}\right)\,
(\varphi')^2\,(|z_1|^2 {-} |z_2|^2)^2 {+}
4M\beta\varphi'\left(1{-}\cos\beta\varphi\right)
\,[1{-}2(|z_|^2 {-} |z_2|^2)]
\,(|z_1|^2 {-} |z_2|^2)\Bigr\}
\end{array} 
\label{pmmp}
\eea
for the second terms in the numerator and denominator of (\ref{vectorratio}).

The next step is to perform the $x$ integration over the
classical soliton configuration, which results in
\bea
\begin{array}{l}
\int dx \,{\cal I}(x,z_1,z_2) = \MAKESPACE\\
\ds
={1\over 2\pi^2} |z_1|^4|z_2|^2
\Biggl\{{64M^2\over 3\mu}[1-2(|z_1^2 - |z_2|^2)]^2
 -{32M^2\over 3\mu}\,[1-4(|z_1|^2 - |z_2|^2)]^2  \MAKESPACE\\
 \phantom{aa}
+ (8\mu)(|z_1|^2 - |z_2|^2)^2 + (4M)(2\pi)[1-2(|z_1|^2 - |z_2|^2)]
 (|z_1|^2 -
|z_2|^2)\Biggr\} \MAKESPACE\\
\ds
={\mu\over 2\pi^2}|z_1|^4|z_2|^2
\Biggl\{{1\over 6}\left({\pi\over N_c}\right)^2 +
N_c\left({\pi\over N_c}\right)^2(|z_1|^2 - |z_z|^2) 
\MAKESPACE\\
 \phantom{aa}
\ds
+\left[8 - \left({\pi\over N_c}\right)^2\left({4\over 3} +
2N_c\right)\right]\left(|z_1|^2 - |z_2|^2\right)^2\Biggr\}
\end{array} 
\label{xInt}
\eea
Specializing to the choices $N_f = N_c = 3$, we find:
\beq
\int dx \,{\cal I}(x,z_1,z_2)= {\mu\over 2\pi^2}|z_1|^4|z_2|^2 
\left\{{\pi^2\over 54} + {\pi^2\over 3} (|z_1|^2 - |z_2|^2)  + \right.  
\left. \left(8 - {22\pi^2\over 27}\right)(|z_1|^2 -|z_2|^2)^2\right\}
\label{final33}
\eeq
This may be evaluated using elementary integrals listed in the Appendix,
yielding
\beq
\langle (\bar u u - \bar d d)^2 \rangle  = 
{\mu\over 7}\left[-{4\over 27} + \left({2\over\pi}\right)^2\right]
\eeq
Substituting this and (\ref{ubddbu}) into (\ref{vectorratio}), 
we obtain our final result
\beq
{T_{20}\over T_{00}} \, = \, 0.54
\eeq
for the vector-current case.

\section{Comments and Discussion}

We have shown in this paper that the matrix elements of  
$\Delta I = 2$ four-fermion operators in QCD$_2$ are not suppressed 
greatly by
comparison with the corresponding $\Delta I = 0$
operators. This conclusion holds true for both scalar-scalar and
vector-vector four-fermion operators. The calculation of the latter case
required the development of some formal machinery, including
collective-coordinate
quantization and the treatment of the time dependence of the quantized
soliton, that had not been required for previous calculations in
QCD$_2$~\cite{FK}.

What might be the significance of our results for the interpretation
of the $\Delta I = 1/2$ enhancement observed in QCD$_4$? Clearly,
QCD$_2$ and QCD$_4$ differ in many respects, particularly in the
ultraviolet and the infrared. The asymptotic freedom of QCD$_4$
yields logarithmic factors in the ultraviolet region, which are
known to leading and next-to-leading order~\cite{pert}. Evaluating these
factors at any plausible renormalization scale, and multiplying
them by the ratios of $T_{00} / T_{20}$ that we find does not
give anything like the $\Delta I = 1/2$ enhancement factor that is
found in QCD$_4$, even if one allows for Clebsch-Gordan factors of order
unity.

What about the infrared features of QCD$_2$ and QCD$_4$? Both theories are
known to have ${\bar q} q$ condensates, and calculations in
QCD$_2$~\cite{FK} of ratios of the
different \hbox{$\langle B|{\bar q} q|B\rangle$}
are known to be in qualitative agreement with
determinations based on the magnitude of the $\sigma$ term extracted from
$\pi N$ scattering~\cite{ChengLi}, and with four-dimensional Skyrme model
calculations~\cite{DN}.

The feeble enhancement of the $\Delta I = 0$ four-fermion operators
is to be compared with the model-dependent result found in the
four-dimensional Skyrme model~\cite{Skyrmehalf}. Might the
problem of the small enhancement that we find
lie with the fact that the gluon condensate is relatively
small in two dimensions, even vanishing in the large-$N_c$
limit and in
the versions of the Skyrme model used in~\cite{Skyrmehalf}? It has indeed
been
suggested that the large enhancement of the $\Delta I = 1/2$ operator
matrix elements in QCD$_4$ might be due to gluon condensate
effects~\cite{glue}.
Our results are certainly consistent with this idea, though by no means
conclusive. How could this suggestion be tested more directly? One
possibility might be to use a formulation of chiral soliton models in
QCD$_4$ in which gluon condensation effects are taken more into account.

Even though our present calculations are not conclusive for the
resolution of the long-standing $\Delta I = 1/2$ puzzle, we believe
that they may help build up a conceptual framework in which it might
be resolved. Some of the calculational formalism developed here may also
be useful in future applications of QCD$_2$.

\bigskip
\vbox{
\begin{flushleft}
{\Large\bf Acknowledgements}
\end{flushleft}

This research was supported by the Israel Science Foundation
administered by the Israel Academy of Sciences and Humanities.
The research of M.K.
was supported in part by
the Einstein Center at the Weizmann Institute and by
a Grant from the G.I.F., the
German-Israeli Foundation for Scientific Research and
Development.}

\vspace{1cm}
\appendix
\noindent
{\Large{\bf Appendix}}
\bigskip
\setcounter{equation}{0}
\def\theequation{A.\arabic{equation}}

In this appendix we collect some formulae useful at intermediate stages
in the derivation of our results for the vector currents.

Some relevant $x$ integrals include:
\beq
\int_{-\infty}^{\infty}\left( \cos \beta \varphi - 1 \right) d x
={-}{4\over\mu}
\label{Icx}
\eeq
and
\beq
\int_{-\infty}^{\infty}\left( \cos \beta \varphi - 1 \right)^2 d x
={16\over3\mu}
\label{IcIIx}
\eeq
used in the scalar-density calculation in section 2, and
\beq
\begin{array}{ccl}
\ds\int dx \left(1 - \cos\beta\varphi\right)^2
& = & \ds {16\over 3\mu} \\
&&\\
\ds\int dx \left(\sin \beta \varphi\right)^2 & = & \ds {8\over 3\mu} \\
&&\\
\ds\int dx \left(\beta \varphi'\right)^2 & = & 8\,\mu
\end{array}
\eeq
used in the vector-current calculation in section 4. For completeness,
we also recall the relation between the two mass scales we have in the
problem
\beq
M  =  {m\over 2\sqrt 2}\left({\pi\over N_c}\right)^{3/2} = \left({\pi\over
 8N_c}\right)
\mu
\eeq
Generic integrals over soliton wave functions are given by
expressions of the general form
\be
\langle |z_i|^{2N}|z_j|^{2P}\rangle =\left\{{(N+n_i)!(P+n_j)!
\over [(N_f+N_c)+(N+P)-1]!}\right\} \bigg /
\left\{{n_i!n_j!\over [(N_f+N_c) - 1]!}\right\}(i\ne j)
\label{genericintegral}
\ee
which is the matrix element of
$|z_i|^{2N}|z_j|^{2P}$ in a normalized state of the form\break
$z_1^{n_1}\,\ldots\,z_i^{n_i}\,\ldots\,z_{N_f}^{n_{N_f}}$\ ,
with the constraint $\sum_i\, n_i = N_c$. The latter
constraint is a quantum consistency condition
imposed by the Wess-Zumino term in the effective
bosonic action, whose coefficient is $N_c$.
To evaluate the matrix elements represented in (\ref{VEVqbqqbq})
we use
\beq
\begin{array}{lcc}
\langle|z_i|^2\rangle & = & \ds {\ds n_i+1\over \ds N_c + N_f}\\
&&\\
\langle|z_i|^2|z_j|^2\rangle & = & \ds {\ds (n_i+1)(n_j+1)
\over \ds (N_c+N_f)(N_c+N_f+1)} (i\ne j)\\
&&\\
\langle|z_i|^4 \rangle & = & \ds 
{\ds (n_i+1)(n_i+2)\over\ds (N_c+N_f)(N_c+N_f+1)}
\end{array}
\label{quadri}
\eeq
in particular. In the case
$N_f=N_c=3$, and normalizing to 1 when $N=2$ and $P=1$,
as is appropriate for the normalization of the $\Delta^+$ state,
we have the general formula
\beq
\langle |z_1|^{2N}\,|z_2|^{2P} \rangle
={5!\over 2} \,{N! \,P!\over (N+P+2)!}
\label{z2N2P}
\eeq
used in sections 2 and 4.

\def\journal#1&#2(#3){\unskip, {\em #1}{\bf\ignorespaces #2}\unskip(#3)}


\begin{thebibliography}{99}
 
\bibitem{thooft}
G. 't Hooft,
{\em Nucl. Phys.} {\bf B72}(1974)461, {\em Nucl. Phys.}{\bf B75}(1974)461.

\bibitem{FS}
Y.~Frishman and J.~Sonnenschein, {\em Phys. Rep.} {\bf 223}(1993)309.

\bibitem{quarks}
J. Ellis, Y. Frishman and M. Karliner,
{\em Phys. Lett.} {\bf B272}(1991)333;
J. Ellis, Y.~Frishman, A. Hanany and M. Karliner,
{\em Nucl. Phys.} {\bf B382}(1992)189.

\bibitem{FK}
Y. Frishman and M. Karliner, {\em Nucl. Phys.} {\bf B344}(1990)393.

\bibitem{DN}
J. Donoghue and C. Nappi,
{\em Phys. Lett.} {\bf B168}(1986)105.

\bibitem{ChengLi}
T.P. Cheng \journal Phys. Rev. &D13(1976)2161;
J. Gasser \etal
\journal Phys. Lett. &B213(1988)85;
M.E. Sainio,
{\em Update of the $\sigma$ term},
Helsinki U. report
HU-TFT-95-36, Jul 1995, invited talk at {\em Sixth International
Symposium on Meson -- Nucleon Physics and Structure of the Nucleon}, 
Blaubeuren, Germany, 10-14 Jul. 1995. 

\bibitem{Erice}J. Ellis and M. Karliner, 
{\em The Strange Spin of the Nucleon}, 
invited lectures at the {\em Int. School of Nucleon Spin Structure},
Erice, August 1995;
CERN-TH/95-334, TAUP-2316-96, hep-ph/9601280.


\bibitem{Cheng} H.-Y. Cheng, {\em Int. J. Mod. Phys} {\bf A4}(1989)495.

\bibitem{pert}G. Buchalla, A.J. Buras and M.E. Lautenbacher,
SLAC-PUB-95-7009, hep-ph/9512380, submitted to {\em Rev. Mod. Phys.}.

\bibitem{Skyrme}
G. Adkins, C. Nappi and \hbox{E. Witten}
 \journal Nucl. Phys. &B228(1983)433; for $N_f{=}3$,
 E. Guadagnini \journal Nucl. Phys. &236(1984)35;
 P.O.~Mazur, M.A.~Nowak and \hbox{M. Prasza\l owicz},
{\em Phys. Lett.} {\bf147B}(1984)137.

\bibitem{glue}
V. Antonelli, S. Bertolini, M. Fabbrichesi, and E.I. Lashin,
{\em Nucl. Phys.} {\bf B469}(1996)181.

\bibitem{Skyrmehalf} For a partial list of references, see
J. Bijnens, H. Sonoda and M.B. Wise, {\em Phys. Lett.} {\bf 140B} (1984)
421;
M. Praszalowicz and J. Trampetic, {\em Phys. Lett.} (1985) 169;
N. Toyota and K. Fujii, {\em Prog. Theor. Phys.} {\bf 75} (1986) 340;
N. Toyota, {\em Prog. Theor. Phys.} {\bf 77} (1987) 688;
K. Fujii, Y. Kondo and S. Saito, {\em Prog. Theor. Phys. Suppl.} {\bf 102}
(1992) 99;

 
\bibitem{Bourquin}
 M. Bourquin {\em et. al.}, {\em Nucl. Phys.} {\bf B241}(1984)1.

 
\end{thebibliography}
\end{document}